\documentclass{optica-article}

\journal{opticajournal}

\articletype{Research Article}

\usepackage{lineno}

\begin{document}


\title{Optical scattering imaging with sub-nanometer precision based on position-ultra-sensitive giant Lamb shift}

\author{Zeyang Liao,\authormark{1,\#} Yuwei Lu,\authormark{1,\#} and Xue-Hua Wang\authormark{1,*}}

\address{\authormark{1}State Key Laboratory of Optoelectronic Materials and Technologies, School of Physics, Sun Yat-sen University, Guangzhou, 510275, Peoples Republic of China }

\homepage{\authormark{\#}These authors contributing equally} 
\email{\authormark{*}wangxueh@mail.sysu.edu.cn} 


\begin{abstract}
The Lamb shift of a quantum emitter very close to a plasmonic nanostructure, mainly induced by the higher-order plasmonic dark modes, can be three or more orders of magnitude larger than that in the free space and it is ultra-sensitive to the emitter position and polarization. We show that this giant Lamb shift can be sensitively observed from the scattering spectrum dip shift of coupled system when the plasmonic nanoparticle or tip scans through the emitter. Based on these observations, we propose an optical localization and polarization microscopy scheme with sub-nanometer precision for a quantum emitter via detecting the scattering spectrum instead of fluorescence. Our method is free of fluorescence quenching problem and it is relatively easier to be implemented in the plasmon-emitter coupling system. Moreover, the sample in our method does not need to be placed inside a plasmonic picocavity to enhance the radiative fluorescence rate and it also works even if the quantum emitter is slightly below a dielectric surface which can bring about broader applications in various fields, such as physics, chemistry, medicine, life science and materials science.
\end{abstract}

\section{Introduction}\label{sec1}

The optical microscope is widely used to localize and image objects which cannot be directly observed by the human eyes. However, the resolution of the conventional optical microscope is subjected to the well-known Abbe's diffraction limit, i.e., the resolvable smallest distance is about half wavelength of the detection light \cite{Vangindertael2018}.  For visible light, the spatial resolution is about 250-300 nm. In the past few decades, several superresolution methods have been proposed to surpass the diffraction limit, such as the stimulated emission depletion microscopy (STED) \cite{Hell2000,Chen2015}, structured illumination microscopy (SIM) \cite{Gustafsson2000,Mudry2012,Zeng2014,Xi2019}, the single-molecule localization microscopy (SMLM) \cite{Betzig2006,Zhuang2006}, and stochastic optical fluctuation imaging (SOFI) \cite{Dertinger2009}. These methods have been widely used for biological imaging with typical resolution being about 20-50 nm. The near-field scanning optical microscope (NSOM) can collect the evanescent field and is in principle not subjected to the diffraction limit, but the achievable minimum resolution to date is about 10 nm in practice \cite{Vobornik2008,Gerton2004,Mauser2014}. With the help of quantum effects, such as using quantum entanglement \cite{Boto2000,Shih2001, Moreau2019}, quantum Rabi oscillations \cite{Liao2010,Liao2012,Rui2016}, quantum statistical imaging \cite{Cui2013,Classen2017, Bhusal2022}, and spatial mode demultiplexing techniques \cite{Tsang2016,Tham2017,Zhou2019}, the optical diffraction limit can also be overcome, but the resolutions demonstrated by these methods are still far from a few nanometers.  To achieve atomic-level resolution, the scanning tunneling microscopy (STM) and atomic force microscopy (AFM) are commonly used \cite{STM1982,AFM2003,Hapala2014}. The resolution gains of STM and AFM are, however, at the expense of losing the spectroscopic properties of the sample and the requirement of extreme environment conditions. Using the highly localized tunneling electrons as a source of excitation, a technique called STM-induced luminescence (STML) can achieve spectral characterization with sub-nanometer spatial resolution \cite{Berndt1993,Chen2010,Zhang2016,Zhang2017}. However, since the electronic tunneling excitation has some limitations, it is a long-standing pursuing goal to achieve sub-nanometer resolution using all-optical methods. 

It is well-known that the plasmonic system can confine the light field in an extremely tiny volume \cite{Goncalves2020,Schuller2010,Chen2018}, which is widely used to enhance the interaction between light and quantum emitters (atom,molecule,quantum dot and so on) \cite{Delga2014,Torma2014,Peng2017,Zeng2012,Andersen2011,Santhosh2016,Liu2017,Hakami2014,Chikkaraddy2016,Zengin2015,Rousseaux2020}. The typical STM junction is actually a plasmonic picocavity which can confine the light field down to sub-nanometer scale \cite{Benz2016,Barbry2015} and it can also enhance the field strength orders of magnitude which has been used for demonstrating the tip-enhanced Raman spectroscopy (TERS) with atomic resolution \cite{Zhang2013,He2019-JACS,Lee2019,He2019-SA}. In addition, the plasmonic picocavity can act as an antenna which can enhance the fluorescence emission rate of a quantum emitter inside the cavity \cite{Kinkhabwala2009,Russell2012,Simovski2020,Baghramyan2022} and it was also applied to demonstrate the tip-enhanced photoluminescence (TEPL) with sub-nanometer resolution \cite{Hou2020}. In both the TERS and TEPL, the target emitters need to be placed inside the plasmonic picocavity which may limit its applications. Considering that the scattering spectrum is much easier to be observed in the plasmonic-emitter coupling system \cite{Tame2013,Liu2021}, it is thus very interesting to devise a plasmonic scanning localization microscopy with ultra-high resolution via its scattering spectrum.  

In this article, we propose an optical scattering imaging (OSI) method with sub-nanometer resolution based on the scattering spectrum shift induced by the plasmon-enhanced Lamb shift when a plasmonic nanoparticle or tip scans through a quantum emitter. As is known,  quantum theory predicts that vacuum is not truly empty, but full of fluctuations where virtual particles are constantly created and annihilated \cite{Scully2010} which can lead to many interesting quantum effects. Among the observable effects of the electromagnetic vacuum, the Lamb shift (LS) is one of the most important effects which directly stimulated the emerging of modern quantum electrodynamics theory, and its precious measurement becomes an important testbed for quantum field theory \cite{Scully2010,Brune1994}. Through vacuum engineering, the LS of a quantum emitter can be significantly modified \cite{Fragner2008,John1990,Marrocco1998,Rentrop2016,Silveri2019,Wang2019,Zhu2000}, e.g., a QE near band edge of photonic crystal can have energy shift one to two orders of magnitudes larger than that in the normal vacuum \cite{Henkel1998,Liu2010,Wang2004}. Due to the much stronger field confinement, the Lamb shift of an emitter very close to a plasmonic nanostructure can be strongly amplified \cite{Sun2008,Vlack2012,Roslawska2022} and has also been experimentally observed \cite{Zhang2017,Hou2020}. Here, we show that the giant Lamb shift in the plasmon-emitter coupling system, mainly induced by the higher-order plasmonic dark modes when the emitter is very close to the plasmonic nanostructure, is ultra-sensitive to the emitter position and it can be observed from the scattering spectrum dip shift of the coupled system with variation of several meV/nm when the metal nanoparticle or tip scans through the quantum emitter.  Due to these observations, we propose that this quantum effect can be exploited to localize the position of an emitter precisely and construct the Lamb shift imaging by scanning the plasmonic nanoparticle or tip through the emitter. The obtained imaging spot is of angstrom size and its shape evidently depends on the dipole orientation of the emitter. Since our method is based on scattering spectrum instead of fluorescence, it is free of fluorescence quenching problem and compared with the usual STML and TEPL methods, our scheme does not require that the sample is placed inside a plasmonic picocavity. The findings here can in principle be developed as a Lamb-shift-based superresolution scanning optical microscope with atomic-level resolution and it also works even when the emitter is embedded slightly below a dielectric substrate,  which can bring about broader applications in various fields.

The article is organized as follows. In Sec. \ref{sec2}, we first describe the model under study and illustrate the theory used to calculate the emission and scattering spectra of this system including the effect of Lamb shift. Particularly, we study the scattering spectrum shift as a function of emitter-nanoparticle distance and propose an experimentally feasible method to observe the giant Lamb shift in this system. In Sec. \ref{sec3}, we propose a possible experimental scheme based on tip-scattering method to detect the giant Lamb shift which can be used to localize the position of an emitter and its polarization even if it is embedded inside a substrate with ultra-high sensitivity. Finally, we summarize our results.

\section{Model and theory}\label{sec2}

\begin{figure}
\centering
\includegraphics[width=0.95\linewidth]{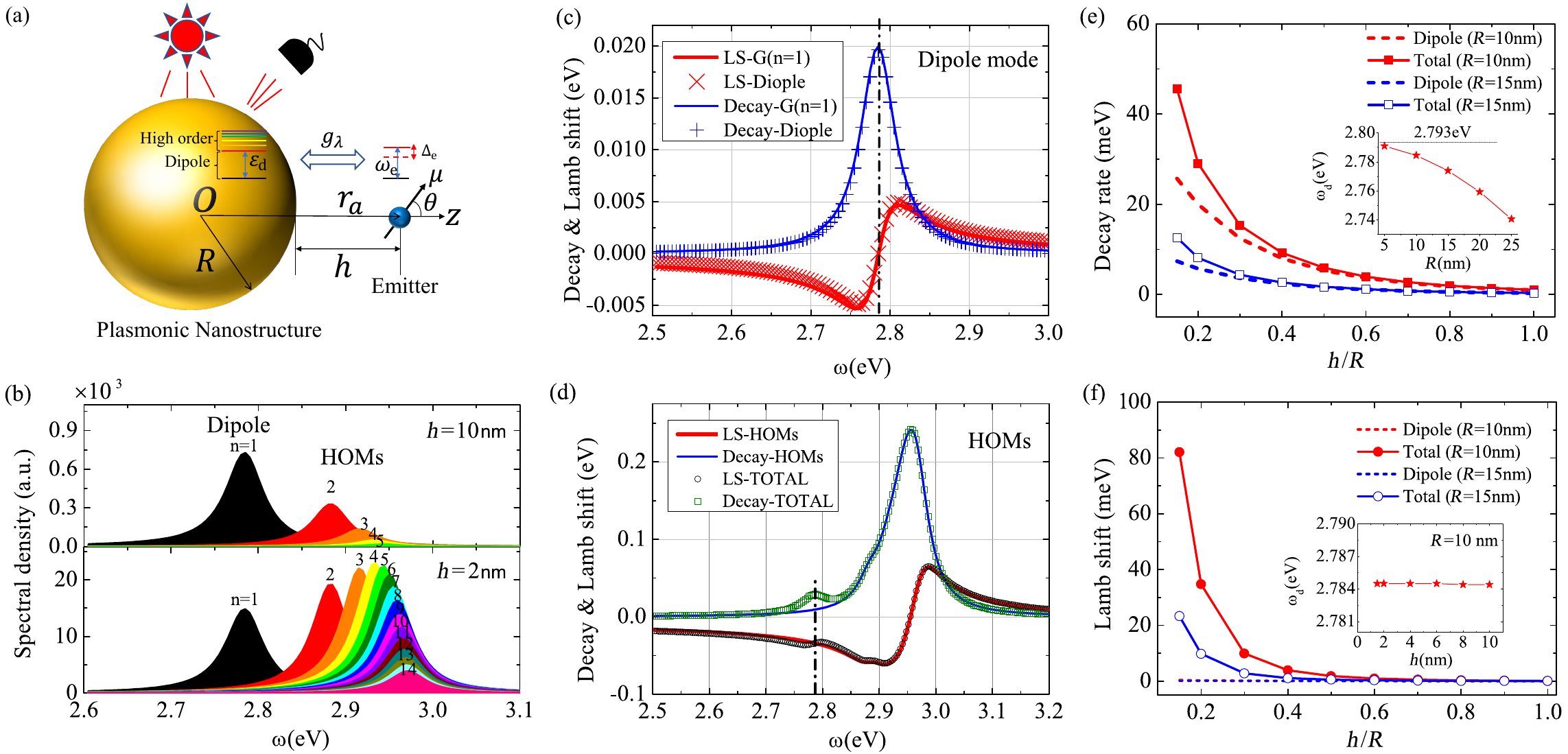}
\caption{(a) QE couples to a nanoparticle. (b) The spectral densities of different plasmonic modes with two different separations (upper panel: $h=10$ nm and lower panel: $h=2$ nm). (c) The spontaneous decay rate and LS of the QE induced by the dipole plasmonic mode. The solid lines are the results calculated by the scattering Green function with $n=1$ and the symbols are those calculated by the effective dipole model. (d) The decay rate (blue thinner line) and Lamb shift (red thicker line) of the QE due to higher-order modes. The total decay rate (green squares) and the total LS (black circles) are also shown.  The dashed dotted lines in both (c) and (d) mark the plasmon dipole frequency when $R=10$ nm and $h=2$ nm. The decay rate (e) and Lamb shift (f) of quantum emitter as a function of $h/R$ with two different nanoparticle radius (i.e., $R=10$ nm and $R=15$ nm) where the curve with symbols are the results considering all orders of plasmonic modes while the dashed curves are those in the dipolar approximation. The insets of (e) and (f) show the plasmonic dipolar frequency as a function of $R$ and $h$, respectively. }
\label{fig1}
\end{figure}

The schematic system considered here is shown in Fig. \ref{fig1}(a) where a QE interacts with a MNP with radius $R$ and  $\mathbf{r}_a$ is the position of the QE with distance $h$ away from the surface of the MNP. In this paper, we mainly consider that $h$ is larger than $1nm$ where the electron tunnelling effect and nonlocal optical response can be neglected \cite{Esteban2012,Raza2015}.  The MNP can support many localized surface plasmon modes with the dipole mode being the bright mode and the higher-order modes (HOMs) being the dark modes \cite{Gong2014,Rousseaux2020}. Considering that the dipole mode can be effectively excited by the incident light and be scattered into the far field while the HOMs do not \cite{Manjavacas2011}, it is convenient to treat the dipole mode as a quantized pseudo-mode separately \cite{Waks2010,Hughes2018,Franke2019} and leave the HOMs as background reservoir fields which can be well described by the macroscopic quantization method based on dyadic Green's function  \cite{Gruner1996,Dung2003}. The effective Hamiltonian is then given by
\begin{equation}\label{eq1}
H=\hbar\left(\omega_{e}-i \frac{\gamma_{e}}{2}\right) \hat{\sigma}^{+} \hat{\sigma}^{-}+\hbar\left(\omega_{d}-i \frac{\gamma_{d}}{2}\right) \hat{d}^{+} \hat{d}+\hbar g_{d e}\left(\hat{\sigma}^{+} \hat{d}+\hat{d}^{+} \hat{\sigma}^{-}\right)+H_{h i},
\end{equation}
where the first term is the effective emitter energy, the second term is the effective plasmon dipole energy, the third term is the interaction between the emitter and the dipole mode, and the four term desribes the effect of the HOMs with
\begin{equation}\label{eq2}
H_{h i}=\hbar \int d \boldsymbol{r} \int d \omega_{\lambda} \omega_{\lambda} \hat{f}^{\prime+}\left(\boldsymbol{r}, \omega_{\lambda}\right) \hat{f}^{\prime}\left(\boldsymbol{r}, \omega_{\lambda}\right)+\left[\hat{\sigma}^{+}+\hat{\sigma}^{-}\right] \boldsymbol{\mu}_{e} \cdot \widehat{\boldsymbol{E}}^{\prime}\left(\boldsymbol{r}_{e}\right).
\end{equation}
The first term in Eq. (2) is the Hamiltonian of the HOMs and the second term is the interaction between the HOMs and QE where the non-rotating interaction terms are retained for the correct calculation of Lamb shift. $\hat{\sigma}^{+}\left(\hat{\sigma}^{-}\right)$ is the Pauli lowering (raising) operator of the QE with transition frequency $\omega_e$, and $\gamma_{e}=\gamma_{e}^{0}+\gamma_{e}^{n r}$ is the emitter decay rate including the radiative part $\gamma_e^0$ and nonradiative part $\gamma_e^{nr}$; $\hat{d}^{+}(\hat{d})$ is the creation (annihilation) operator of the plasmon dipole mode with frequency $\omega_d$ and linewidth $\gamma_d$ which is the combination of radiative decay $\gamma_{d}^{0}$ and nonradiative decay $\gamma_{d}^{nr}$; $g_{d e}=\omega_{\mathrm{d}}^{2} \mu_{0} \boldsymbol{\mu}_{e} \cdot \boldsymbol{G}_{0}\left(\boldsymbol{r}_{e}, \boldsymbol{r}_{d} ; \omega_{d}\right) \cdot \boldsymbol{\mu}_{d} / \hbar$ is the coupling strength between the emitter and the plasmon dipole mode, here $\mathbf{\mu_{e}}$ and $\mathbf{\mu_{d}}$ are the dipole moments of the QE and the nanoparticle, respectively. Here, we assume that $|\mathbf{\mu_{e}}|=24D$ \cite{Savasta2010} and the magnitude of the effective plasmon dipole moment $\left|\boldsymbol{\mu}_{d}\right|=\epsilon_{b} \sqrt{12 \pi \epsilon_{0} \hbar \eta_{1} R^{3}}$ where $\eta_1 = \left\{\left.\frac{d}{d \omega} R e\left[\epsilon_{m}(\omega)\right]\right|_{\omega=\omega_{d}}\right\}^{-1}$ and $\epsilon_{m}(\omega)=\epsilon_{\infty}-\omega_{p}^{2} /\left(\omega^{2}+i \omega \gamma_{p}\right)$ is the effective permittivity of the nanoparticle in the Drude model with $\epsilon_b$ being the relative permittivity of the surrounding environment and $\epsilon_0$ is the vacuum permittivity \cite{Ridolfo2010}. In all the numerical calculations throughout this paper, we choose typical parameters for silver with $\epsilon_\infty=6.0$, $\omega_p=7.9$eV, $\gamma_p=51$meV and  $\gamma_e=15$meV \cite{Vlack2012}.  Since the free vacuum and the dipolar field have been accounted separately, $\hat{f}^{\prime+}\left(r, \omega_{\lambda}\right)\left(\hat{f}^{\prime}\left(r, \omega_{\lambda}\right)\right)$ denotes the continuum bosonic-field creation (annihilation) operator of the HOMs and $\widehat{\boldsymbol{E}}^{\prime}(\boldsymbol{r})=\widehat{\boldsymbol{E}}^{\prime+}(\boldsymbol{r})+\widehat{\boldsymbol{E}}^{\prime-}(\boldsymbol{r})$ is the electric field operator of the HOMs with
\begin{equation}\label{eq3}
\hat{E}^{\prime+}\left(\boldsymbol{r}_{e}\right)=i \sqrt{\frac{\hbar}{\pi \epsilon_{0}}} \int \frac{\omega_{\lambda}^{2}}{c^{2}} d \omega_{\lambda} \int d \boldsymbol{r}^{\prime} \sqrt{\epsilon_{I}\left(\boldsymbol{r}^{\prime}, \omega_{\lambda}\right)} \sum_{n \geq 2}^{\infty} \mathbf{G}_{n}^{S}\left(\boldsymbol{r}, \boldsymbol{r}^{\prime} ; \omega_{\lambda}\right) \cdot \hat{f}^{\prime}\left(\boldsymbol{r}^{\prime}, \omega_{\lambda}\right)  
\end{equation}
being the positive part of the field where $\epsilon_{I}\left(\boldsymbol{r}^{\prime}, \omega_{\lambda}\right)$ is the imaginary part of the material permittivity and $\mathbf{G}_{n \geq 2}^{\mathbf{S}}\left(\boldsymbol{r}, \boldsymbol{r}^{\prime} ; \omega_{\lambda}\right)$ is the higher-order scattering Green function with $\mathrm{n} \geq 2$. The dyadic Green function $\boldsymbol{G}\left(\boldsymbol{r}, \boldsymbol{r}^{\prime} ; \omega_{\lambda}\right)$ satisfies the equation
\begin{equation}\label{eq4}
\boldsymbol{\nabla} \times \boldsymbol{\nabla} \times \mathbf{G}\left(\boldsymbol{r}, \boldsymbol{r}^{\prime} ; \omega_{\lambda}\right)-\frac{\omega_{\lambda}^{2}}{c^{2}} \epsilon\left(\boldsymbol{r}, \omega_{\lambda}\right) \mathbf{G}\left(\boldsymbol{r}, \boldsymbol{r}^{\prime} ; \omega_{\lambda}\right)=\mathbf{I} \delta\left(\boldsymbol{r}-\boldsymbol{r}^{\prime}\right).
\end{equation}
Without plasmonic structure, $\boldsymbol{G}\left(\boldsymbol{r}, \boldsymbol{r}^{\prime} ; \omega\right)=\boldsymbol{G}_{0}\left(\boldsymbol{r}, \boldsymbol{r}^{\prime} ; \omega\right)$. In the presence of plasmonic structure, $\boldsymbol{G}\left(\boldsymbol{r}, \boldsymbol{r}^{\prime} ; \omega\right)=\boldsymbol{G}_{0}\left(\boldsymbol{r}, \boldsymbol{r}^{\prime} ; \omega\right)+\boldsymbol{G}^{S}\left(\boldsymbol{r}, \boldsymbol{r}^{\prime} ; \omega\right)$ where $\boldsymbol{G}_{0}\left(\boldsymbol{r}, \boldsymbol{r'} ; \omega\right)$ and $\boldsymbol{G}^{S}\left(\boldsymbol{r}, \boldsymbol{r}^{\prime} ; \omega\right)$ are the free and scattering parts of the dyadic Green function, respectively and their analytical expressions are given in the Supplementary Information (SI) Sec. I. The normalized spectral density of the nth-order plasmonic mode at position $\mathbf{r}$ is given by $\mathrm{J}_{\mathrm{n}}(\boldsymbol{r}, \omega)=\operatorname{Im}\left[\hat{\boldsymbol{e}}_{\boldsymbol{i}} \cdot \boldsymbol{G}_{n}^{S}(\boldsymbol{r}, \boldsymbol{r} ; \omega) \cdot \hat{\boldsymbol{e}}_{\boldsymbol{i}}\right] / G_{0}$ where $\boldsymbol{G}_{n}^{S}(\boldsymbol{r}, \boldsymbol{r} ; \omega)$ is the nth-order scattering Green function of the nanoparticle, $G_0=k/6\pi$ and $\hat{\boldsymbol{e}}_{\boldsymbol{i}}$ is a unit vector \cite{Vlack2012}. When the distance between the emitter and the nanoparticle is of the order of or larger than the radius of the nanoparticle, the dipole mode dominates (see upper panel of Fig. \ref{fig1}(b)). However, when the emitter is very close to the nanoparticle, the HOMs can have larger spectral density than that of the dipole mode (see lower panel of Fig. \ref{fig1}(b)) and therefore they should be taken into accounts. 

\begin{figure}[t]
\centering
\includegraphics[width=0.95\linewidth]{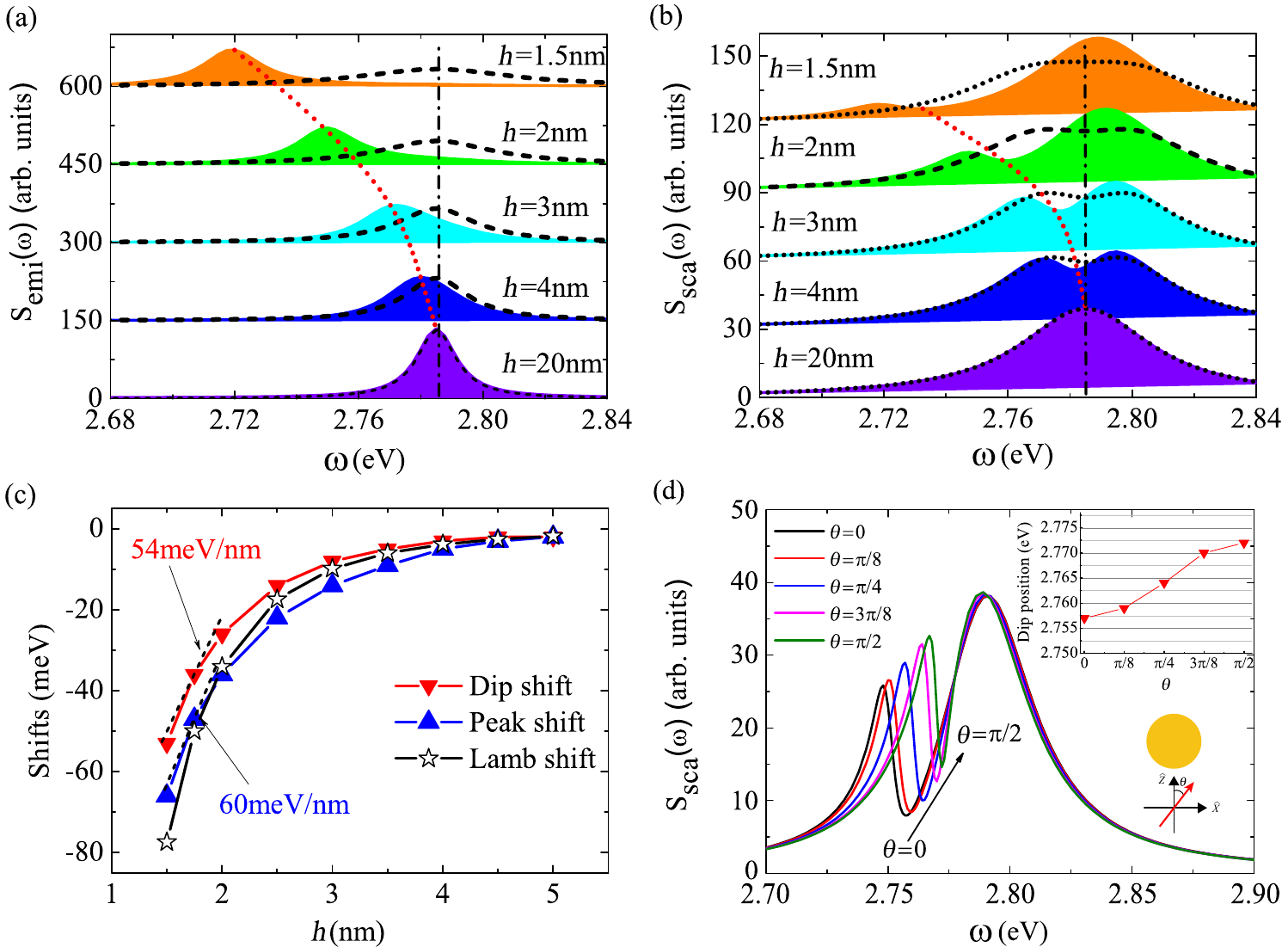}
\caption{The emission (a) and scattering (b) spectrum of the coupled system for different emitter-nanoparticle distances (the curves with filling area). The black dotted curves are the results without Lamb shift. In both (a) and (b), the emitter is assumed to be along the direction of dipole polarization. (c) The Lamb shift, scattering spectrum dip shift, and emission peak shift as a function of distance. (d) The scattering spectrum changes with different polarization orientations of the emitter with $h=2$ nm. Inset: the scattering spectrum dip shift as a function of $\theta$. $R=10$ nm and $\omega_e=\omega_d=2.785$ eV.}
\label{fig2}
\end{figure}

For spontaneous decay process where only the emitter is initially excited, we can obtain the emission spectrum of the emitter given by (see SI Sec. II)
\begin{equation}\label{eq5}
S_{e m i}(\omega) \propto\left|\omega-\omega_{e}^{\prime}-\frac{g_{d e}^{2}}{\omega-\omega_{d}^{\prime}}\right|^{-2}.
\end{equation}
Here, we define the effective transition frequency of the quantum emitter $\omega_{e}^{\prime}=\omega_{e}+\Delta_{e}^{\prime}(\omega)-i\left[\gamma_{e}+\gamma_{\mathrm{e}}^{\prime}(\omega)\right] / 2$ in which
\begin{equation}
\Delta_{e}^{\prime}(\omega)=-\frac{\omega^{2}}{\hbar \epsilon_{0} c^{2}} \boldsymbol{\mu}_{\mathbf{e}} \cdot \operatorname{Re} \sum_{n \geq 2} \mathbf{G}_{n}^{S}\left(\boldsymbol{r}_{a}, \boldsymbol{r}_{a} ; \omega\right) \cdot \boldsymbol{\mu}_{e}
\end{equation}
and 
\begin{equation}
\gamma_{\mathrm{e}}^{\prime}(\omega)=\frac{2 \omega^{2}}{\hbar \epsilon_{\mathrm{o}} c^{2}} \boldsymbol{\mu}_{e} \cdot \operatorname{Im} \sum_{n \geq 2} \mathbf{G}_{n}^{S}\left(\boldsymbol{r}_{a}, \boldsymbol{r}_{a} ; \omega\right) \cdot \boldsymbol{\mu}_{e}
\end{equation}
are the Lamb shift and decay rate induced by the HOMs vacuum fields, respectively. The term $g_{d e}^{2} /\left(\omega-\omega_{d}^{\prime}\right)$ with $\omega_{d}^{\prime}=\omega_{d}-i \gamma_{d} / 2$ is the contribution from the plasmon dipole mode.  The total LS of the emitter is given by $\Delta_{e}^{\prime}(\omega)+\operatorname{Re}\left[g_{d e}^{2} /\left(\omega-\omega_{d}^{\prime}\right)\right]$ and the total decay rate is given by $\gamma_{e}+\gamma_{\mathrm{e}}^{\prime}(\omega)-2 \operatorname{Im}\left[g_{d e}^{2} /\left(\omega-\omega_{d}^{\prime}\right)\right]$. The emission spectrum shown in Equation (\ref{eq5}) is actually equivalent to that obtained from the original Hamiltonian of the system in the rotating wave approximation, i.e. $S_{e m i}(\omega)=\left\{\left[\omega-\omega_{e}-\Delta_{e}(\omega)\right]^{2}+\Gamma_{e}^{2}(\omega) / 4\right\}^{-1}$ \cite{Wang2003} where $\Delta_{e}(\omega)=-\frac{\omega^{2}}{\hbar \epsilon_{0} c^{2}} \boldsymbol{\mu}_{e} \cdot \operatorname{Re}\left[\boldsymbol{G}^{S}\left(\boldsymbol{r}_{e}, \boldsymbol{r}_{e} ; \omega\right)\right] \cdot \boldsymbol{\mu}_{e}$ and $\Gamma_{e}(\omega)=\frac{2 \omega^{2}}{\hbar \epsilon_{0} c^{2}} \boldsymbol{\mu}_{e} \cdot \operatorname{Im}\left[\boldsymbol{G}^{S}\left(\boldsymbol{r}_{e}, \boldsymbol{r}_{e} ; \omega\right)\right] \cdot \boldsymbol{\mu}_{e}$  are the LS and the effective decay rate calculated by the full scattering Green function. The equivalence can be seen from Fig. \ref{fig1}(c) where the well agreement of both the real and imaginary parts of $g_{d e}^{2} /\left(\omega-\omega_{d}^{\prime}\right)$ with $-\left(\omega^{2} / \hbar \epsilon_{0} c^{2}\right) \boldsymbol{\mu}_{e} \cdot \boldsymbol{G}_{n=1}^{S}\left(\boldsymbol{r}_{e}, \boldsymbol{r}_{e} ; \omega\right) \cdot \boldsymbol{\mu}_{e}$ are shown (also see SI Sec.IV). Around the dipolar mode frequency $\omega_d$, the LS induced by the dipole mode vanishes but the decay rate is maximum. On the contrary, the HOMs have little effect on the decay rate but have significant effect on the LS even though the bare emitter transition frequency is far-off resonant from the HOMs’ frequencies, as shown in Fig. \ref{fig1}(d). The LS induced by the HOMs can be high up to 34.1 meV in current example which is more than three orders of magnitude larger than that in the free vacuum. The decay rate and Lamb shift of the quantum emitter as a function of $h/R$ for two different nanoparticle radii (i.e., $R=10$ nm and $R=15$ nm) are shown in Figs. \ref{fig1}(e) and (f), respectively. For comparisons, the results in the dipolar approximations are also shown as the dashed curves. We can see that both the total decay rate and the Lamb shift increase rapidly when $h/R$ decreases and their values especially the Lamb shift when $h/R$ is small can be very different from those in the dipole approximation. We also find that both the decay rate and the Lamb shift increase when $R$ decreases for the same $h/R$.  By extracting the peak of the density of state of the dipolar component, we find that the plasmon dipolar frequency does not depend on $h$ but is red-shifted when $R$ increases (see the inset of Fig. \ref{fig1}(e) and Fig. \ref{fig1}(f)). In the small radius limit, the dipolar frequency approaches the value of $\omega_p/\sqrt{\epsilon_\infty+2\epsilon_b}$ in the quasistatic-limit which is $2.793$ eV in our example. Since $\boldsymbol{G}^{S}\left(\boldsymbol{r}_{e}, \boldsymbol{r}_{e} ; \omega\right)$ strongly depends on the position of the emitter, the LS of the emitter varies with emitter’s position. From Fig. 2(a), we can see that the emission spectrum peak is red-shifted and the linewidth is also broaden when the nanoparticle moves towards the emitter which provides a method to observe the LS in this system and similar phenomena have also been experimentally observed in the plasmonic nanocavity system \cite{Hou2020}. Although the coupling strength increases when the emitter is closer to the nanoparticle, we do not observe fluorescence splitting because the damping rate also increases rapidly and thus no spectrum splitting is observed in Fig. \ref{fig2}(a).  We should also mention that in Fig. \ref{fig2}(a) we show the emitter fluorescence spectrum shape but its real intensity in the far field also depends on the quantum yield. In this coupling system, since the emitter radiative decay $\gamma_{e}^{0}$ is usually much less than its non-radiative decay rate $\gamma_{e}^{nr}$  especially when the quenching effect is considered in the case that the emitter is very close to the metal nanoparticle \cite{Hu2021}, the quantum yield of the fluorescence is actually very small and most of the emitted energy are absorbed by the metal nanoparticle. In addition,  the spectrum shown in Equation (\ref{eq5}) is the fluorescence spectrum when only the emitter is initially excited which is not easy to be realized in this system because the cross section of the nanoparticle is usually much larger than that of the quantum emitter. To enhance the quantum yield of the fluorescence, a plasmonic cavity together with a dielectric spacer is usually required \cite{Hou2020} which may limit its application scenarios. Next, we show that the LS can also be observed from the scattering spectrum of the system under weak driving field, which is much easier to be experimentally implemented.  

According to the input-output theory \cite{Gardiner1985}, the output field operator $\hat{a}_{out}(t)=\sqrt{\gamma_{d}^{0}}\hat{d}(t)+\sqrt{\gamma_{e}^{0}}\sigma^{-}(t)$ and the scattering light spectrum $S(\omega)\propto \langle \hat{a}^{+}_{out}(\omega)\hat{a}_{out}(\omega)\rangle$ where $\hat{a}_{out}(\omega)$ is the Fourier transformation of $\hat{a}_{out}(t)$. In the typical nanoparticle-emitter coupling system, $\gamma_{d}^{0}\gg \gamma_{e}^{0}$ and therefore the output field is mainly due to the emission of the palsmon dipole, i.e., $\hat{a}_{out}(t)\approx\sqrt{\gamma_{d}^{0}}\hat{d}(t)$  \cite{Hu2021}. Thus, the scattering spectrum of the system in the stationary limit $\mathrm{S}_{\mathrm{sca}}(\omega) \propto \lim _{t \rightarrow \infty} \operatorname{Re}\left[\int_{0}^{\infty}\left\langle d^{+}(t) \mathrm{d}(t+\tau)\right\rangle e^{i \omega \tau} d t\right]$ and in the weak excitation limit is given by (see Sec. III in SM) 
\begin{equation}\label{eq8}
S_{s c a}(\omega) \propto-\operatorname{Im}\left[\omega-\omega_{d}^{\prime}-\frac{g_{d e}^{2}}{\left(\omega-\omega_{e}^{\prime}\right)}\right]^{-1}.
\end{equation}
If the coupling strength vanishes (i.e., $g_{de}=0$), the scattering spectrum $S_{s c a}(\omega) \propto-\operatorname{Im}\left(\omega-\omega_{d}^{\prime}\right)^{-1}$ which has Lorentzian lineshape with frequency $\omega_d$ and linewidth $\gamma_d$. Without coupling, the scattering spectrum of the nanoparticle does not contain any information of the emitter. However,  if $g_{de}\neq 0$, the effective transition frequency of the plasmon dipole is modified in the presence of the QE and hence $S_{sca} (\omega)$ includes the emitter’s spectral information.  The scattering spectrum shown in Equation (\ref{eq8}) can be rewritten as
\begin{equation}\label{eq9}
S_{s c a}(\omega) \propto-\operatorname{Im}\left[\frac{f_{+}}{\omega-\omega_{+}}+\frac{f_{-}}{\omega-\omega_{-}}\right],
\end{equation}
where $\omega_{\pm}=\frac{1}{2}\left(\omega_{e}^{\prime}+\omega_{d}^{\prime} \pm \Delta_{l s}\right)$ are two eigenfrequencies of the coupled system, with $\Delta_{l s}=\sqrt{\left(\omega_{d}^{\prime}-\omega_{e}^{\prime}\right)^{2}+4 g_{d e}^{2}}$ and $\mathrm{f}_{\pm}=\frac{1}{2} \pm \frac{\omega_{d}^{\prime}-\omega_{e}^{\prime}}{\Delta_{l s}}$ are two constant coefficients \cite{Liu2021}. It is clearly seen that the scattering spectrum is the superposition of two eigen-channels and quantum interference between these two channels can induce a spectrum dip even if the system is in the pseudo-strong coupling regime. If $f_+=f_-$ and they are real, the spectrum dip occurs at exactly the center of the two eigen-frequencies, i.e., $\frac{1}{2}\left(\omega_{e}+\Delta_{e}^{\prime}+\omega_{d}\right)$ which is a linear function of the LS. However, in the usual case, $f_+ \neq f_-$ and both of them may be complex number, the spectrum dip usually deviates from $\frac{1}{2}\left(\omega_{e}+\Delta_{e}^{\prime}+\omega_{d}\right)$ but it is still a monotonic function of the emitter LS (see SI Sec. V).

The scattering spectrum as a function of the emitter distance is shown in Fig. \ref{fig2}(b). Different from the spontaneous emission spectrum, there is usually a spectrum dip in the scattering spectrum due to the Fano-like interfernce even if the strong coupling condition is not met \cite{Zhang2017,Liu2021}. From Fig. \ref{fig2}(b) we can see that the LS can be observed from the shift of the spectrum dip when the nanoparticle moves towards the emitter (the curves with filling area). In contrast, without considering the LS, the position of spectrum dip does not change with the emitter position (dotted curves) although the peak separation increases with decreasing distance. The LS, emission peak shift and scattering spectrum dip shift as a function of emitter position are shown in Fig. \ref{fig2}(c). It is clearly seen that both the emission peak and the scattering spectrum dip are rapidly red-shifted together with the LS  when $h$ decreases and the gradients are about 50$\sim$60 meV/nm when $h\approx 2$ nm. Therefore, either the emission peak shift or the scattering dip shift can reveal the LS of the emitter. The method discussed above also works when the emitter transition frequency is slightly different from the plasmon dipole frequency (see SI Sec. VI).

Since the LS highly depends on the relative position between the emitter and the nanoparticle, it is possible to determine an emitter’s position from the scattering spectrum. From Fig. \ref{fig2}(c) we can see that when the distance decreases from 20 nm to 1.5 nm, the magnitude of LS increases from almost 0 to 77.4 meV which can be observed from the scattering spectrum dip shift of about 53 meV. In particular, when the emitter-nanoparticle distance changes from 2 nm to 1.5 nm, the magnitude of LS increases by about 48 meV and the spectrum dip is red-shifted by about 27 meV. In a typical high-resolution spectrometer, 0.02 nm wavelength difference (about 0.1 meV energy shift) can be resolved. Thus, from the scattering spectrum dip shift we can in principle measure emitter position change with angstrom or even sub-angstrom precision which is only limited by the precision position control of the nanoparticle. The precision of the scheme may be further improved if the emitter-nanoparticle distance is less than 1 nm, while in this regime the quantum effects such as quantum tunneling and nonlocal effects  may play an important role which needs to be considered \cite{Karanikolas2021,Esteban2012,Raza2015,Goncalves2020, Cirac2019,Zhou2022}.

In addition to the distance, the polarization direction of the quantum emitter can also affect the LS and therefore the scattering spectrum is changed as the emitter polarization angle.  In the near-field regime, $\mathbf{G}_{0}\left(\boldsymbol{r}_{e}, \boldsymbol{r}_{d} ; \omega\right) \approx e^{i k r}\left(-\boldsymbol{I}+3 \mathbf{e}_{\mathrm{r}} \mathbf{e}_{\mathrm{r}}\right) / 4 \pi k^{2} r^{3}$ where $\mathrm{r}=\left|\boldsymbol{r}_{e}-\boldsymbol{r}_{d}\right|$. The coupling strength $g_{\mathrm{de}} \approx \mu_{d} \mu_{e} \sqrt{1+3 \cos ^{2} \theta} / 4 \pi \epsilon_{0} \hbar(R+h)^{3}$ which decreases when $\theta$ increases from 0 to $\pi$/2. In addition to the coupling strength, the magnitude of LS also decreases when $\theta$ increases from 0 to $\pi$/2 (see SI Sec. VII). The scattering spectra for different emitter polarizations are shown in Fig. \ref{fig2}(d) where it is shown that the spectrum dip is shifted when $\theta$ varies. The dip shift is maximum when the polarization of the emitter is along the z-direction while it is minimum when it is along the x-direction (Fig. \ref{fig2}(d)). The spectrum dip shift with respect to $\theta$ is also shown in the inset of Fig. \ref{fig2}(d). The average gradient of the shift is about 0.17 meV per degree. Hence it is possible to detect the polarization of the emitter with about $1^{\text{o}}$ resolution from the scattering spectrum. 

 \begin{figure}[t]
\centering
\includegraphics[width=\linewidth]{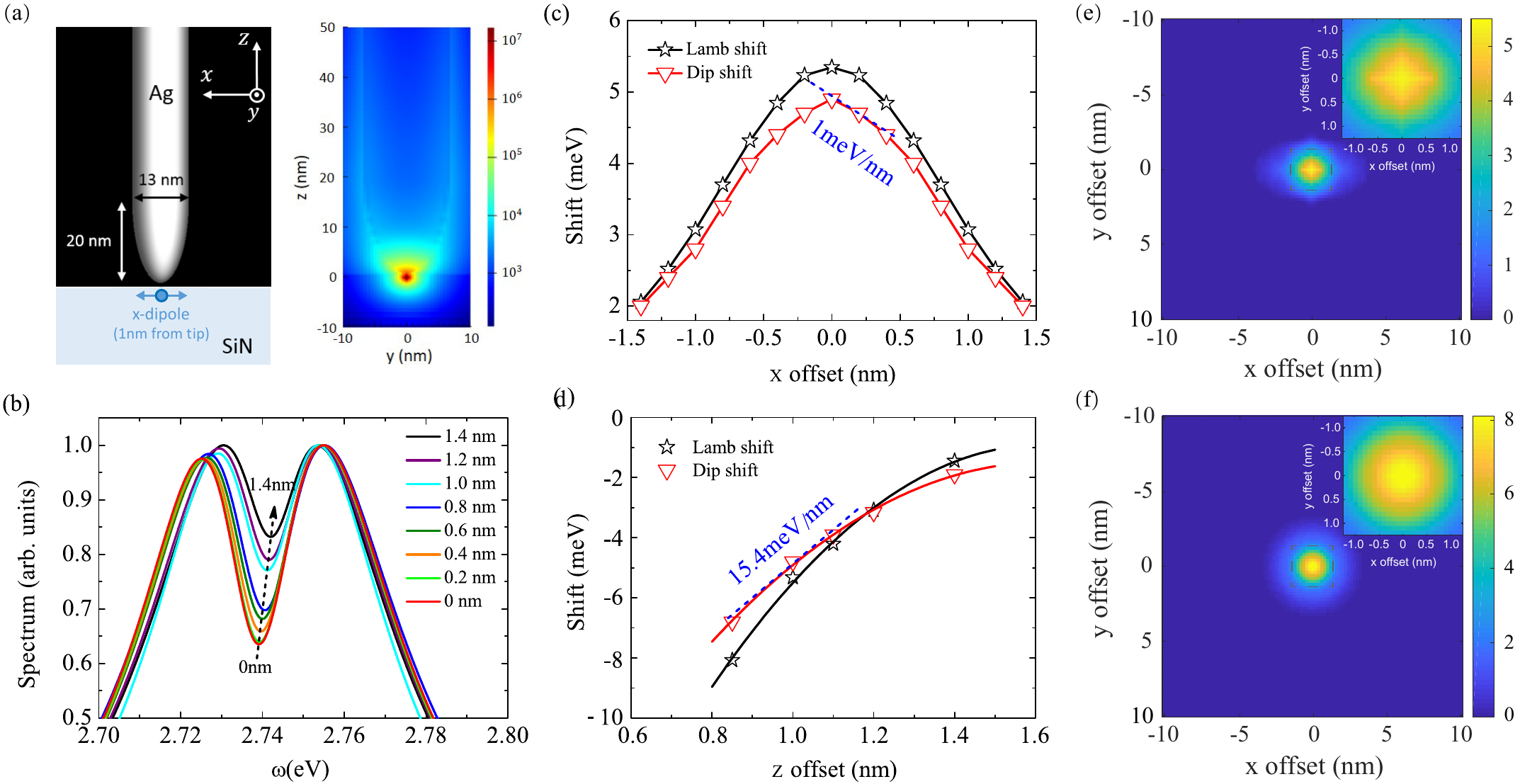}
\caption{(a) A schematic setup for ultrasensitive emitter localization based on tip plasmonics induced LS (left) and the field intensity distribution around the tip when the tip is 1 nm above the emitter (right). The tip has a semi elliptical shape and the major axis is 20 nm and minor axis is 13 nm. (b) The scattering spectrum of the coupled plasmonic system for different lateral offsets ($0\rightarrow 1.4 $ nm) of the tip positions when the z-offset is fixed to be 1 nm. (c,d) The LS (black stars) and the spectrum dip shift (red down triangles) for different lateral tip positions with 1 nm z-offset (c) and for different $z$ offset when the tip is right on top of the emitter (d). (e,f) The extracted 2D image of the emitter based on the dip shift of the scanning scattering spectrum when the emitter is x-polarized (e) or z-polarized (f) when the z-offset is 1 nm, where the color represents the amount of the spectrum dip shift. In all these figures, the emitter is assumed to be 0.5 nm below the SiN substrate surface.}
\label{fig3}
\end{figure} 

\section{Experiment proposal}\label{sec3}

Finally, we propose a possible realization based on the theory described above to detect the position of a quantum emitter sitting on or inside a substrate (such as SiN) as shown in Fig. \ref{fig3}(a). Here, as an example the emitter is assumed to be x-polarized and 0.5 nm below the surface. The field intensity distribution when the tip is 0.5 nm above the surface is also shown in right panel of Fig. \ref{fig3}(a), where we can see that the field is strongly localized around the tip. A light beam is applied to the tip and the scattering spectra for different tip lateral positions are shown in Fig. \ref{fig3}(b) when the z distance between the emitter and tip (z-offset) is fixed to be 1 nm. When the tip is closer to the emitter, the spectrum dip is shifted to the lower frequency due to the LS discussed above. The dip shift and the LS as a function of lateral offset are shown in Fig. \ref{fig3}(c) when the z-offset is fixed to be 1 nm. When the tip is right on top of the emitter, the spectrum dip has a maximum shift. The gradient around the maximum shift is about 1 meV/nm near the center.  When the tip is right on top of the emitter, the spectrum dip shift as a function of z distance is shown in Fig. \ref{fig3}(d). When the tip moves towards the emitter, the spectrum dip is red-shifted with gradient about 15.4 meV/nm, which indicates a much higher sensitivity for detecting the longitudinal distance. For a spectrometer with 0.02 nm resolution, the lateral and longitudinal resolutions in this setup can in principle be about 1 $\mathring{A}$  and 0.1 $\mathring{A}$, respectively. If the substrate is replaced by a material with lower reflective index or the emitter is placed above the substrate surface, the LS can be even larger and the resolution can be further enhanced (see SI Sec. VIII). We can also extract a two-dimensional image of the emitter from the dip shift of the scattering spectra at different positions when the emitter is x-polarized (Fig. \ref{fig3}(e)) or z-polarized (Fig. \ref{fig3}(f)). When the emitter is z-polarized, the image is symmetric. However, when the emitter is x-polarized, the image has an elliptical shape with major axis along the polarization direction. Thus, in our scheme, we can not only detect the position of the emitter but also its polarization direction. When the z-offset increases, the major axis of the imaging spot also increases which indicates the reduction of resolution (see Fig. S6 in SM). Different from the typical STML and TEPL methods \cite{Berndt1993,Chen2010,Zhang2016,Hou2020}, here our scheme does not require the formation of plasmonic picocavity to enhance the fluorescence rate  and  it also works even when the emitter is embedded slightly below a dielectric substrate which may find more application scenarios.

\section{Summary and discussion}\label{sec4}
To conclude, we show that a quantum emitter very close to a metal nanoparticle or tip can have huge Lamb shift mainly induced by the higher-order plasmonic dark modes and this energy shift is ultra-sensitive to the emitter position. We also show that this giant Lamb shift can be sensitively observed from the scattering spectrum dip shift when the metal nanoparticle or tip scans through the emitter. Moreover, we propose that this quantum effect can be exploited for all-optical detection of an emitter position with angstrom precision which is comparable to the typical STM, STML and TEPL methods and we can also determine the polarization direction of the quantum emitter. In contrast to the typical STML and TEPL methods which are usually surface bound and only detect the sample in a plasmonic picocavity, we can detect the position of the emitter precisely even if it is slightly below a dielectric substrate. Due to the fact that the experimental measurement of the scattering spectrum is much easier than the fluorescence in the plasmon-emitter coupling system, our work here can find important applications in many areas and can stimulate extensive theoretical and experimental researches in the future. Finally, we should mention that the emitter is treated as a point dipole in this work to demonstrate the basic principle, the full scattering spectrum considering the non-dipolar effect of the quantum emitter and other quantum effects should be further studied in the future \cite{Lyu2022}.  


\medskip

\begin{backmatter}
\bmsection{Funding}
\noindent The authors thank R. Liu and X. Zeng for helpful discussions. This work was supported by the National Key R\&D Program of China (Grant No. 2021YFA1400800),  the Key-Area Research and Development Program of Guangdong Province (Grant No.2018B030329001), the Guangdong Special Support Program (Grant No.2019JC05X397), the Natural Science Foundations of Guangdong (Grant Nos. 2021A1515010039 and 2018A030313722).


\bmsection{Disclosures}
\noindent The authors declare no conflicts of interest.

\bmsection{Data availability} Data underlying the results presented in this paper are not publicly available at this time but may be obtained from the authors upon reasonable request.

\bmsection{Supplemental document}
See Supplement 1 for supporting content. 

\end{backmatter}




\end{document}